\documentclass[reprint,amsmath,amssymb,aps,pra]{revtex4-2}
\usepackage[backref=none]{hyperref}
\usepackage{graphicx}
\usepackage{dcolumn}
\usepackage{comment}
\usepackage{lipsum}
\usepackage{xcolor}
\usepackage{tikz}
\usepackage{physics}
\usepackage{bm}
\usepackage{soul}
\usetikzlibrary{arrows.meta,positioning}
\begin{document}

\preprint{APS/123-QED}

\title{Time-independent counterdiabatic driving for emergent two-level subspaces in many-body systems}
\author{Simon Dengis$^{1}$}
\email{simon.dengis@lkb.upmc.fr}
\author{Peter Schlagheck$^{2}$}
\email{peter.schlagheck@uliege.be}

\affiliation{
 $^{1}$Laboratoire Kastler Brossel, Sorbonne Université, CNRS, ENS-Universit\'e PSL, Collège de France, 4 Place Jussieu, 75005 Paris, France}
\affiliation{
  $^{2}$CESAM research unit, University of Liege, B-4000 Li\`ege, Belgium}

\begin{abstract}
We show that geodesic motion in the Riemannian manifold of quantum states provides a direct route to time-independent counterdiabatic driving. Using the relation between the counterdiabatic Hamiltonian and the quantum metric tensor, we prove that a constant-speed geodesic makes the Hilbert-Schmidt norm of the counterdiabatic Hamiltonian constant. For effective two-level systems whose counterdiabatic correction has a fixed operator direction, this further implies that the full counterdiabatic Hamiltonian itself is time independent. We illustrate this result with the Landau-Zener model, three-level Stimulated Raman adiabatic passage and a collectively driven Rydberg ensemble in the blockade regime. Limitations of this approach in realistic many-body systems are discussed, where the two-level reduction is only emergent and leakage out of the effective subspace bounds the achievable speedup. In all cases, time-independent counterdiabatic driving achieves unit-fidelity state preparation on timescales substantially shorter than conventional adiabatic protocols while replacing temporally shaped auxiliary controls by fixed-amplitude fields.

\end{abstract}

\maketitle
\textit{Introduction.} Adiabatic state preparation provides a robust route to high-fidelity quantum control, as a system initially prepared in an instantaneous eigenstate remains locked to this eigenstate when the Hamiltonian is varied sufficiently slowly \cite{Born_1928,Landau_1932,Zener_1932, Stuck_1932,Majorana_1932, Berry_1987,Unanyan_1997}. This robustness, however, comes at the price of long protocol durations, which are often incompatible with decoherence, losses, or technical noise. Shortcuts to adiabaticity were developed to overcome this limitation by reproducing the outcome of an adiabatic evolution in a shorter and finite time \cite{Chen_2010,Bason_2012,delCampo_2013,Ban_2014, Deffner_2014,delCampo_2019, Chen_2021}.
Among these methods, counterdiabatic driving aims to identify the precise terms responsible for diabatic transitions during the evolution and to compensate for them exactly \cite{Berry_2009,Demirplak_2003}. While this method has been extensively developed in the paradigmatic case of few-level systems \cite{Fleischhauer_1999, Chen_2010,Kolodrubetz_2017,Guery-Odelin_2019}, its implementation in many-body systems is considerably more challenging due to the requirement of knowing the instantaneous spectrum of the system throughout the protocol. Even though a number of approximation schemes have been developed to circumvent this limitation \cite{Sels_2017,Claeys_2019, Prielinger_2021, Mbeng_2022, Petiziol2024}, the resulting counterdiabatic terms generally require non-trivial time-dependent control fields whose experimental implementation may be particularly demanding.
\begin{figure}[!t]
    \centering
    \includegraphics[width=0.45\textwidth]
    {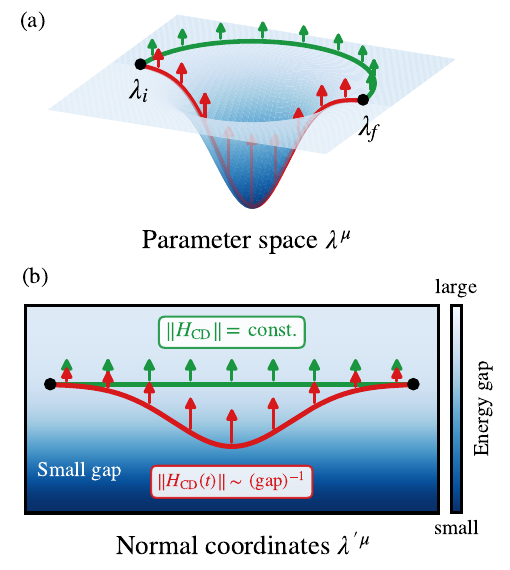}
    \caption{\textbf{Geometric origin of constant counterdiabatic driving}. Geodesic protocols correspond to straight lines in Riemann normal coordinates, yielding a constant quantum speed and therefore a constant norm of the counterdiabatic Hamiltonian. In contrast, arbitrary protocols experience a parameter-dependent quantum metric, causing the required counterdiabatic field to become stronger in regions where the energy gap is small. The two panels illustrate the equivalent representation obtained after transforming to normal coordinates, where the geometry is flattened while the gap structure remains encoded in the background. Arrow lengths represent the counterdiabatic Hamiltonian norm, which is time-independent for geodesic parametrization.}
    \label{fig:illustration}
\end{figure}

Recent studies have shown that, within this framework, it is possible to define a counterdiabatic Hamiltonian whose norm is constant \cite{Dengis2025_a, Dengis2025_b, Balducci2026}. This remarkable property arises from selecting the optimal adiabatic driving protocol for the system, defined by the geodesics of the parameter space \cite{Carlini_2006,Tomka_2016, Kolodrubetz_2017, Nauth_2022}. Since this space forms a Riemannian manifold \cite{Provost_1980, Hetenyi2023}, it is naturally endowed with a metric that allows one to identify the parametric trajectories maximizing, at every instant, the fidelity between the evolving state and the target state. When the system follows such a trajectory, its energy variance, which is linked to the Hilbert-Schmidt norm of the counterdiabatic Hamiltonian \cite{Kolodrubetz_2017} remains constant throughout the evolution \cite{Misner1973,Tomka_2016, Thesis_dengis}. This shortcut-to-adiabaticity protocol has recently enabled proposals of experimentally feasible schemes for NOON state creation with ultracold atoms \cite{Dengis2025_a, Dengis2025_b} and the adiabatic crossing of quantum phase transitions \cite{Balducci2026}.

Here we show that the time dependence of the counterdiabatic Hamiltonian can itself be removed for a large class of systems. For effective two-level systems whose counterdiabatic correction points along a fixed direction in operator space, this implies that every matrix element of the counterdiabatic Hamiltonian becomes time independent. We illustrate this result with the Landau-Zener model and with the three-level Stimulated Raman adiabatic passage (STIRAP) protocol. We further show that this result extends to many-body systems exhibiting an emergent two-level reduction, illustrated here with a collectively driven ensemble of Rydberg atoms in the blockade regime.

\emph{Theoretical framework.}
We consider a system described by a Hamiltonian $H(\lambda^{\mu}(t))$ that is time dependent through an ensemble of parameters labeled $\lambda^\mu\equiv(\lambda^{(1)},\lambda^{(2)},...)$. For an eigenstate $\ket{n(\lambda^\mu)}$ of the Hamiltonian, the distance between two infinitesimally separated points of parameter space defines a natural Riemannian metric on this manifold. The corresponding line element reads
\begin{equation}
ds^2 = 1-\big\vert n(\lambda^\mu+d\lambda^\mu)-
n(\lambda^\mu)\big\vert^2
= g_{\mu\nu}^{(n)}\,d\lambda^\mu d\lambda^\nu,
\end{equation}
where 
\begin{equation}
g_{\mu\nu}^{(n)}
=
\mathrm{Re}\big\{\langle \partial_\mu n|\partial_\nu n\rangle
-
\langle \partial_\mu n|n\rangle
\langle n|\partial_\nu n\rangle\big\}
\end{equation}
is the quantum metric tensor related to eigenstate $\vert n\rangle \equiv \vert n(\lambda^\mu)\rangle$, and $\vert \partial_\mu n \rangle \equiv \partial \vert n\rangle /\partial \lambda^\mu $. This quantity determines the local distinguishability of neighboring eigenstates and therefore quantifies how rapidly the quantum state changes under a variation of the control parameters. Defining the length functional 
$L = \int dt \, (g_{\mu\nu}^{(n)}\dot{\lambda}^\mu \dot{\lambda}^\nu)^{1/2}$, the minimization condition $\delta L = 0$ leads to the geodesic equation \cite{Misner1973, Rezakhani2009, Tomka_2016} 
\begin{equation}\label{eq:geodesic}
\ddot{\lambda}^\mu + \Gamma^{\mu}_{\alpha\beta}\dot{\lambda}^\alpha \dot{\lambda}^\beta = 0,
\end{equation}
where $\Gamma^{\mu}_{\alpha\beta}$ are the Christoffel symbols. Curves $\lambda^\mu$ that satisfy Eq.(\ref{eq:geodesic}) are of stationary length and satisfy $g_{\mu\nu}^{(n)}\dot{\lambda}^\mu \dot{\lambda}^\nu=\mathrm{const.}$ In a Riemannian manifold, there always exist the so-called normal coordinates $\lambda'$ centered at an initial point, for which every $\Gamma^{\mu}_{\alpha\beta}$ vanish \cite{Misner1973}. In that case, the geodesic equation reduces to $\ddot{\lambda}'^\mu = 0,$ whose solution is simply $\lambda'^\mu(t) = \lambda'^\mu(0) + \left[\lambda'^\mu(T)-\lambda'^\mu(0)\right]t/T$. 

Beyond being a well-known property of geodesic curves \cite{Misner1973}, this result also provides a practical method for explicitly determining the optimal trajectory connecting two quantum states, often referred to as the quantum brachistochrone \cite{Carlini_2006,Nauth_2022, Balducci2026}. In particular, the quantity $g_{\mu\nu}^{(n)}\dot{\lambda}^\mu \dot{\lambda}^\nu$ is directly related to the Hilbert-Schmidt norm of the counterdiabatic Hamiltonian $H_\mathrm{CD}$ through its diagonal matrix elements, according to \cite{delCampo2012}
\begin{equation}\label{eq:GCD}
    \Vert H_\mathrm{CD}\Vert^2 = \sum_n\langle n\vert H_\mathrm{CD}^2\vert n\rangle  =\hbar^2G_{\mu\nu}\dot{\lambda}^\mu\dot{\lambda}^\nu,
\end{equation}
where $G_{\mu\nu} = \sum_n g_{\mu\nu}^{(n)}$ is the sum of the quantum metric tensors associated with the instantaneous eigenstates of $H$. For a geodesic parametrization (see Supplemental Material), the right-hand side of Eq.(\ref{eq:GCD}) is constant and the norm of the counterdiabatic Hamiltonian is therefore time independent. The apparent time dependence of counterdiabatic driving is, in many relevant cases, a coordinate artifact: when the dynamics is expressed in the natural geodesic coordinate of the quantum metric, the counterdiabatic driving becomes uniform. Since the total cost of the protocol along the evolution is given by $C=\int_0^Tdt\, \Vert H_\mathrm{CD}(t) \Vert$, it is minimized when following a geodesic and is given by $C=\hbar \ell_{geo}$, with the geodesic length $\ell_{geo}$ (see Supplemental Material).

However, a constant norm does not, in general, imply that the full counterdiabatic Hamiltonian is time independent as it only fixes the length of the vector representing it in operator space. The direction of this vector may still rotate during the protocol. 
For a two-level control Hamiltonian $H(t) = \boldsymbol{h}(t)\cdot \boldsymbol{\sigma}$ for which one component of $\boldsymbol{h}$ vanishes for all $t$ (i.e. $h_y(t)=0$), then the direction of the associated counterdiabatic Hamiltonian $H_\mathrm{CD}(t)$ is fixed along the uncontrolled axis ($H_\mathrm{CD}(t) = \Omega_y(t)\,\sigma_y$). If in addition $\boldsymbol{h}(t)$ follows a geodesic of the induced quantum metric at constant speed, then Eq.(\ref{eq:GCD}) fixes $\Omega_y(t)$ to a constant, yielding a counterdiabatic Hamiltonian entirely time independent.


\emph{Landau-Zener systems.} The Landau-Zener model is one of the paradigmatic models for the study of adiabatic quantum dynamics. In particular, it is among the few time-dependent systems for which the time evolution of the state populations can be obtained analytically \cite{Landau_1932, Zener_1932, Stuck_1932, Majorana_1932}. The Hamiltonian describing the Landau-Zener model is given by
\begin{equation}\label{eq:HLZ}
    H_\mathrm{LZ}(t) = \lambda(t)\,\sigma_z + J(t)\,\sigma_x,
\end{equation}
where $\lambda(t)$ and $J(t)$ are the control parameters. The associated counterdiabatic Hamiltonian, expressed in polar coordinates according to $\lambda(t)=r(t)\sin(\theta)$ and $J(t)=r(t)\cos(\theta)$, takes the particularly simple form $H_\mathrm{CD}(t) = -\hbar\dot{\theta}(t)/2\,\sigma_y$ \cite{Berry_2009,delCampo_2013}.
\begin{figure}[!t]
    \centering
    \includegraphics[width=0.475\textwidth]{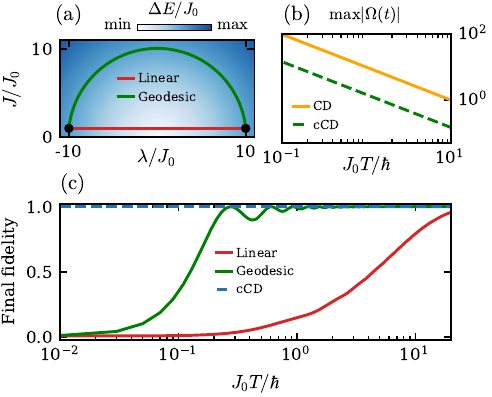}
    \caption{\textbf{Application of constant counterdiabatic driving (cCD) to Landau-Zener system.} \textbf{(a)} Gap landscape of the Landau-Zener Hamiltonian, with $\Delta E = 2\sqrt{\lambda^2 + J^2}$ and $J_0\equiv J(0)$. The geodesic path (green) avoids the small-gap region associated with the avoided crossing that would require longer adiabatic time, in contrast to the conventional linear protocol (red). \textbf{(b)} Maximum counterdiabatic amplitude required to suppress nonadiabatic excitations, as a function of the protocol duration T. The constant counterdiabatic term reduces the cost by one order of magnitude compared to the usual counterdiabatic driving. \textbf{(c)} Final fidelity of the state-transfer protocol. Geodesic driving alone already provides a substantial enhancement over conventional adiabatic evolution, while counterdiabatic driving achieves perfect state preparation. The use of a geodesic driving still allows for unit fidelity while lowering the experimental cost of the protocol.}
    \label{fig:LZ}
\end{figure}
The geometry induced by the time dependence of $H(t)$ is simply that of the Bloch sphere generated by its instantaneous eigenstates. The line element associated with the symmetric space $SU(2)$ is given by $ds^2 = d\theta^2/4$, yielding a quantum metric tensor for which the only non-vanishing element is independent of $\theta$ such that $g_{\theta\theta}=1/4$. The corresponding geodesic for $\theta(t)$ is therefore described by a linear protocol, $\theta(t) = \theta_i + (\theta_f-\theta_i)\,t/T,$ where $\theta_{i,f} = \tan^{-1}(\lambda_{i,f}/J_{i,f})$ are the initial and final angles of the driving protocol. These angles approach $\pm \pi/2$ when $\vert \lambda_{i,f}\vert \gg J_{i,f}$, entering the asymptotic region of the arctangent. The optimal paths in $SU(2)$ correspond to circle trajectories at fixed gap \cite{Tomka_2016}. As a consequence, the traversal speed is constant and directly fixes the amplitude of the counterdiabatic driving, yielding
\begin{equation}
    H_\mathrm{CD} = -\frac{(\theta_f -\theta_i)\hbar}{2T}\,\sigma_y
\end{equation}
Thus, the Hamiltonian $H_\mathrm{LZ}(t)+H_\mathrm{CD}$ drives the system adiabatically along its instantaneous eigenstates without requiring any additional time dependence in the counterdiabatic term. The addition of a constant control field is sufficient to exactly suppress all diabatic transitions.
While the practical implementation of a counterdiabatic Hamiltonian is often challenging due to the need for a control term that is absent from the original Hamiltonian, rendering this term time independent already significantly reduces the experimental complexity associated with such protocols. 

Figure \ref{fig:LZ} illustrates several aspects of constant counterdiabatic driving applied to the Landau-Zener Hamiltonian (\ref{eq:HLZ}). In particular, Fig.~\ref{fig:LZ}(a) displays two possible driving protocols in the parameter space surrounding the energy gap between the two levels, represented by the color gradient. A geodesic protocol maintains a constant gap throughout the entire evolution, thereby enabling adiabatic transport at a constant speed  \cite{Tomka_2016, Balducci2026}. Figure~\ref{fig:LZ}(b) compares the maximum counterdiabatic amplitude $\Omega(t) = \dot{\theta}(t)/2$ required to drive the system adiabatically for usual and time-independent CD, and shows that the latter reduces the required control amplitude by one order of magnitude. Figure~\ref{fig:LZ}(c) presents the final fidelity as a function of the total protocol duration, which remains at unity when counterdiabatic terms are included, even though these terms are completely time independent.

\emph{STIRAP}. Stimulated Raman adiabatic passage describes a protocol for coherent state transfer in three-level quantum systems \cite{Bergmann1998,Chen_2010,Vitanov2017}. It enables the population inversion between two states that are not directly coupled, while avoiding occupation of an intermediate state that may be subject to losses or decoherence \cite{Bergmann1998}. Originally developed in atomic and molecular physics, STIRAP has since found applications in a wide range of platforms, including trapped ions \cite{Feng2008} and ultracold atoms \cite{zhang_2024}. The system is described by the Hamiltonian
\begin{equation}\label{H_STIRAP}
    H_\mathrm{S} =
    \begin{pmatrix}
        0 & \Omega_p(t) & 0 \\
        \Omega_p(t) & \Delta & \Omega_s(t) \\
        0 & \Omega_s(t) & \delta
    \end{pmatrix},
\end{equation}
where $\Omega_p(t)$ and $\Omega_s(t)$ denote the pump and Stokes Rabi frequencies, respectively. The detuning $\Delta$ measures the energy mismatch between the intermediate state and the applied laser frequencies, while $\delta$ corresponds to the two-photon detuning between the initial and target states.
\begin{figure}[!t]
    \centering
    \includegraphics[width=0.475\textwidth]{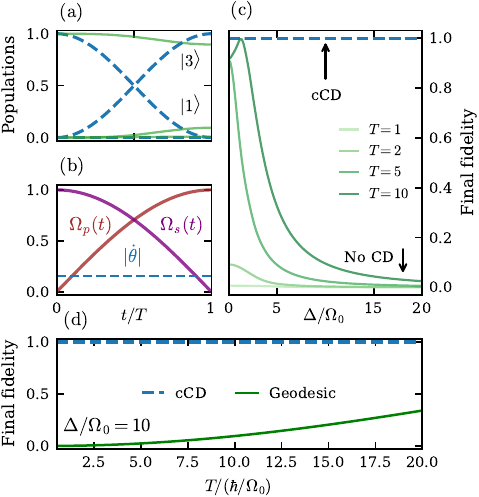}
    \caption{\textbf{Application of constant counterdiabatic driving to STIRAP.} \textbf{(a)} Population transfer from $\vert1\rangle$ to $\vert 3 \rangle$ along the geodesic path, for $T=10\hbar/\Omega_0$. \textbf{(b)} Geodesic pulse sequence generating a constant counterdiabatic correction. \textbf{(c)} Final fidelity versus detuning $\Delta$ for different protocol durations. The constant-amplitude counterdiabatic field completely suppresses the loss of fidelity observed in the standard protocol. \textbf{(d)} Final fidelity versus protocol duration for $\Delta/\Omega_0=10$. Constant counterdiabatic driving achieves perfect state preparation over the entire range of durations considered, demonstrating that the geodesic construction extends naturally beyond effective two-level systems, without any loss of fidelity.}
    \label{fig:STIRAP}
\end{figure}
When both detunings vanish, $\delta=\Delta=0$, one recovers the resonant STIRAP configuration, often referred to as the \emph{single-photon resonance} case. The condition $\Delta \neq \delta=0$ defines the \emph{two-photon resonance} regime, for which the energy difference between the initial and final states exactly matches the frequency difference of the two driving fields. In this case, the Hamiltonian possesses a dark eigenstate whose energy remains exactly zero throughout the evolution. This regime is particularly important since it allows for an exact analytical description of the adiabatic passage and provides a natural framework for implementing shortcut-to-adiabaticity techniques. 

We will focus on the case $\delta=0$. 
Defining the couplings as $\Omega_p(t)=\Omega(t)\sin(\theta(t))$ and $\Omega_s(t)=\Omega(t)\cos(\theta(t)),$ where $\tan(\theta(t))=\Omega_p(t)/\Omega_s(t)$ and $\Omega(t)=\sqrt{\Omega_p^2(t)+\Omega_s^2(t)}$, the Hamiltonian (\ref{H_STIRAP}) can be diagonalized, and its spectrum is well known. Among its eigenstates, the dark state $\vert D(\theta)\rangle$ corresponds to the zero-energy eigenvalue and is given by $\vert D(\theta)\rangle = \cos(\theta)\vert 1\rangle - \sin(\theta)\vert 3\rangle$, whose geometry does not depend on $\Omega(t)$ which we fix to $\Omega_0$. Consequently, adiabatically driving the dark state from the initial value $\theta=0$ to the final value $\theta=\pi/2$ transfers the system from state $\vert 1\rangle$ to state $\vert 3\rangle$. In particular, the geodesic followed by the angle $\theta(t)$ can be obtained by evaluating the only non-vanishing component of the quantum metric tensor associated with the dark state, yielding $g_{\theta\theta}=1$. The geodesic equation can then be solved straightforwardly, leading to a linear evolution of the angle, $\theta(t)=\theta_i+(\theta_f-\theta_i)t/T,$ with $\theta_i=0$ and $\theta_f=\pi/2$. Therefore, the optimal path is characterized by the following parametrization: $\label{theta_geo}
    \theta(t) = \pi t/2T$. 
The matrix elements of the counterdiabatic Hamiltonian associated with the driving of the dark state can be written as follows \cite{Bergmann1998,Vitanov2017}:
\begin{equation}
H_\mathrm{CD}(t) = i\hbar\dot{\theta}(t) \,( \vert 1\rangle \langle 3\vert - \vert 3\rangle \langle 1\vert).
\end{equation}
The use of the constant driving protocol immediately fixes the matrix elements of the counterdiabatic Hamiltonian, yielding $H_\mathrm{CD} = i\pi\hbar/2T\left(\vert 1\rangle\langle 3\vert - \vert 3\rangle\langle 1\vert\right).$ The system can therefore be driven adiabatically using counterdiabatic terms that are entirely time independent. This example shows that the emergence of time-independent counterdiabatic driving is not restricted to two-level Hamiltonians. It also appears whenever the relevant adiabatic eigenstate explores a one-dimensional geodesic manifold and the associated counterdiabatic operator has a fixed direction. The relevant ingredient is therefore not the dimension of the Hilbert space, but the effective dimension of the adiabatic manifold explored by the target eigenstate.

Figure \ref{fig:STIRAP} illustrates constant counterdiabatic driving applied to the STIRAP system. The use of such a protocol enables complete population transfer, and hence unit fidelity of the target state $\vert 3\rangle$, on timescales significantly shorter than those achievable with a purely geodesic protocol (see Fig.~\ref{fig:STIRAP}(d)). The control amplitudes for such a driving are depicted in Fig.~\ref{fig:STIRAP}(b). In Fig.~\ref{fig:STIRAP}(c), we show that in this parameter regime, the final fidelity remains independent of the detuning $\Delta$ when constant counterdiabatic driving is employed, whereas it progressively decreases with increasing $\Delta$ in the absence of counterdiabatic terms.
Finally, Fig.~\ref{fig:STIRAP}(d) demonstrates that unit fidelity is achieved for any protocol duration $T$ when the time-independent counterdiabatic terms are included. Although the explicit time dependence of the counterdiabatic Hamiltonian has been completely eliminated, the efficiency of the population-transfer protocol remains unaffected.

\textit{Rydberg blockade.} The previous examples rely on an exactly two-dimensional (or exactly reducible) adiabatic manifold, which the target eigenstate never leaves. A physically more realistic test of our construction is a genuine many-body system, where the two-level structure is only emergent: the reduction to an effective Landau-Zener subspace holds only approximately, and the exponentially many remaining levels of the Hilbert space may be populated once the counterdiabatic coupling competes with the energy scales protecting the subspace. To probe to what extent time-independent counterdiabatic driving survives in this setting, we consider the preparation of Rydberg $\vert W\rangle$ states \cite{Moller2008,Browaeys2020}, where the blockade interaction V provides such a protective energy scale. The many-body Hamiltonian describing a chain of $N$ two-level Rydberg atoms is given by
\begin{equation}\label{eq:HRYD}
    H_\mathrm{Rb}= -\frac{\Delta(t)}{2}\sum_{i=1}^{N}\sigma^z_i + \frac{\Omega}{2}\sum_{i=1}^{N}\sigma_i^x + \frac{V}{4}\sum_{i<j}^N (1+\sigma_i^z)(1+\sigma_j^z),
\end{equation}
where $\sigma_i^{x,y,z}$ are the Pauli operators of the two-level atom $i$, $\Delta$ is the detuning between the laser frequency and the atomic transition, $\Omega$ is the laser Rabi frequency, and $V$ denotes the Rydberg-Rydberg interaction strength. In the blockade regime $V \gg \Omega$, the dynamics become restricted to a two-dimensional Hilbert space spanned by the ground state $\vert G\rangle$ (in which all atoms remain unexcited) and the singly excited symmetric state $\vert W\rangle$ \cite{Jaksch2000,Lukin2001, Saffman2010}. In this limit, the effective Hamiltonian describing the two-level subspace takes the form $H_\mathrm{eff}=-(\Delta(t)/2)\,\sigma_z + (\sqrt{N}\Omega/2)\,\sigma_x,$ which is strictly equivalent to the single-control Landau-Zener Hamiltonian (\ref{eq:HLZ}) under the substitutions $\lambda(t)\rightarrow -\Delta(t)/2$ and $J\rightarrow \sqrt{N}\Omega/2$. The corresponding geodesic protocol is therefore given by $\Delta(t)=\sqrt{N}\Omega\tan\left[\alpha_i+(\alpha_f-\alpha_i)\,t/T\right],$ where $\alpha_{i,f}=\tan^{-1}(\Delta_{i,f}/\sqrt{N}\Omega)$ with $\Delta_i\equiv\Delta(t=0)$ and $\Delta_f\equiv\Delta(t=T)$. Under this geodesic parametrization, the matrix elements of the counterdiabatic Hamiltonian become time independent and are given by $-i(\alpha_f-\alpha_i)\hbar/2T$ which tend to  $-i\pi\hbar/2T$ for $\vert \Delta_{i,f}/\sqrt{N}\Omega\vert \gg 1$.
A convenient way to embed the counterdiabatic correction into the original many-body Hamiltonian is to introduce a counterdiabatic amplitude $\Omega_\mathrm{CD}=-\pi\hbar/\sqrt{N}T$ \cite{Dengis2025_a, Dengis2025_b}, yielding a complex Rabi frequency $\Omega + i\Omega_\mathrm{CD}$. With this definition, the counterdiabatic contribution is naturally incorporated into the many-body dynamics via a $\sigma^y$ term. The corresponding Hamiltonian reads
\begin{align}\label{eq:HRGCD}
    H_\mathrm{Rb}= -\frac{\Delta(t)}{2}\sum_{i=1}^{N}\sigma^z_i + \frac{1}{2}\sum_{i=1}^{N} \big(\Omega\,\sigma_i^x + \Omega_\mathrm{CD}\,\sigma_i^y \big) \\
    \notag+ \frac{V}{4}\sum_{i<j}^N (1+\sigma_i^z)(1+\sigma_j^z)
\end{align}
The resulting Hamiltonian (\ref{eq:HRGCD}) reproduces the adiabatic evolution prescribed by constant counterdiabatic driving within the complete many-body structure of the Rydberg system. It is worth noting that the complex Rabi frequency can be efficiently emulated by a suitably chosen high-frequency Floquet modulation \cite{Goldman_2014, Petiziol_2018, Dengis2026}.

\begin{figure}[!t]
    \centering
    \includegraphics[width=0.475\textwidth]{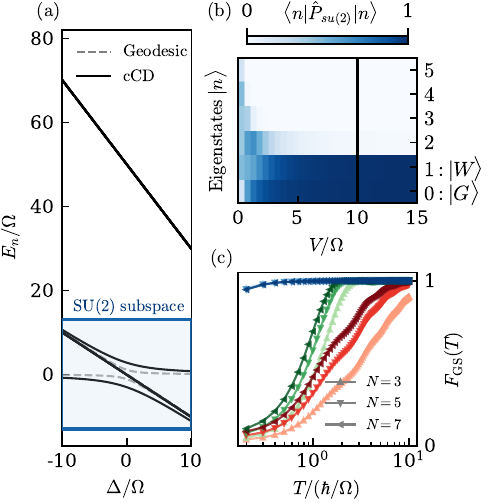}
    \caption{\textbf{Application of constant counterdiabatic driving to a collectively driven Rydberg ensemble in the blockade regime, for $V=50\Omega$}. \textbf{(a)} Many-body spectrum as a function of the laser detuning $\Delta$. The shaded region highlights the effective Landau-Zener subspace spanned by the collective ground state $\vert G\rangle $ and the singly excited state $\vert W\rangle$, in the strong Rydberg interaction regime $V\gg \Omega$, for $N=5$ and $T=0.5\hbar/\Omega$. \textbf{(b)} Overlap between the Landau-Zener subspace projector $\hat{P}_{su(2)}=\vert G\rangle \langle G\vert + \vert W\rangle \langle W\vert$ and the eigenstates of the Hamiltonian (\ref{eq:HRYD}) as a function of the blockade ratio $V/\Omega$, for $\Delta=0$ and $N=5$. \textbf{(c)} Final fidelity $F_\mathrm{GS}(T)$ as a function of the protocol duration for different atom numbers N.}
    \label{fig:RYD}
\end{figure}

Figure \ref{fig:RYD} presents the results obtained with the many-body Hamiltonian (\ref{eq:HRGCD}) for the adiabatic preparation of the $\vert W\rangle$ state. Figure~\ref{fig:RYD}(a) illustrates the effect of the time-independent counterdiabatic term on the many-body spectrum. The Landau-Zener gap occurring between the states $\vert G\rangle$ and $\vert W\rangle$ is effectively enlarged by the action of the counterdiabatic correction.
Figure~\ref{fig:RYD}(b) shows the overlap between the eigenstates of Hamiltonian (\ref{eq:HRYD}) and the subspace of interest composed of the states $\vert G\rangle$ and $\vert W\rangle$. The Rydberg blockade regime allows for the restriction of the relevant Hilbert space to only two states, yielding an effective two-level system in which the constant counterdiabatic driving can be defined. Finally, Fig.~\ref{fig:RYD}(c) compares the final fidelities, $F_\mathrm{GS}(T)=\vert\langle W\vert\psi(T)\rangle\vert^2,$ as a function of the total protocol duration $T$ for different atom numbers. For both the linear and geodesic protocols, the collective enhancement factor $\sqrt{N}$ appearing in the coupling term of $H_\mathrm{eff}$ explains why the adiabatic regime is reached more rapidly as the number of particles increases. In all cases, however, the cCD protocol maintains fidelities close to unity for evolution times that are substantially shorter than those required by conventional adiabatic protocols, despite the complete time independence of the counterdiabatic correction. 

Since the counterdiabatic correction is constructed within the effective two-level subspace, residual leakage appears at very short times, where the required auxiliary coupling becomes large enough to break the blockade approximation. The cCD scales as $\pi\hbar/\sqrt{N}T$, so that shortening the protocol eventually makes it comparable to the interaction scale V that isolates the $\{ \vert G\rangle ,\vert W\rangle\}$ manifold. This sets a minimal protocol duration $T \geq \pi \hbar/\sqrt{N}V$ below which leakage towards doubly excited states becomes significant, as visible at short times in Fig.~\ref{fig:RYD}(c). Moreover, the correction compensates only the diabatic couplings within the emergent subspace: residual effects of order $\Omega/V$ remain uncorrected and would require higher-order or variational counterdiabatic schemes \cite{Claeys_2019, Cepaite_2023, Ohga2026}.

\textit{Conclusion.} In this work, we have shown that the connection between the counterdiabatic Hamiltonian and the quantum metric tensor naturally enables the construction of a $H_\mathrm{CD}$ with constant norm and, in the case of emergent two-level systems, with entirely time-independent matrix elements. This property emerges directly from the use of geodesics in parameter space, which ensure a constant velocity with respect to the quantum metric and thereby transform conventional counterdiabatic driving into a fixed-amplitude protocol.
We have demonstrated that this construction applies not only to the paradigmatic Landau-Zener model, but also to three-level systems such as STIRAP, as well as to many-body systems exhibiting an emergent two-level subspace, exemplified here by Rydberg atom ensembles in the blockade regime. In all cases considered, constant counterdiabatic driving achieves unit-fidelity state preparation on timescales substantially shorter than those required by conventional adiabatic protocols, while simultaneously reducing the experimental complexity associated with the auxiliary control fields.

Our proposal relies on the existence of a parameter regime in which the many-body dynamics reduces to an effective two-level system. Although the counterdiabatic term is constant in time, its amplitude scales as $1/T$: driving the system too fast therefore requires a counterdiabatic field large enough to compete with the energy scale that protects the subspace, so that the dynamics leaks out of the effective two-level manifold and the Landau-Zener reduction no longer holds. Rather than being a limitation, this constant-amplitude protocol provides a baseline for further optimization: it can serve as a natural starting point for quantum optimal control schemes aimed at shortening the preparation time while improving the robustness against leakage out of the subspace.

\textit{Acknowledgments.} We thank Sandro Wimberger, Marin Bukov and Adolfo del Campo for inspiring discussions. This project (EOS 40007526) has received funding from the FWO and F.R.S.-FNRS under the Excellence of Science (EOS) programme. S.D. acknowledges funding from the European Union’s Horizon Europe
Framework Programme (EIC Pathfinder Challenge project Veriqub) under Grant Agreement
No. 101114899.

\bibliography{biblio}

\newpage

\section*{Supplemental material}
\subsection*{Time-independent counterdiabatic Hamiltonian}\label{app:Proof}
We start from the quantum geometric line element \cite{Provost_1980}
\begin{equation}
    ds^2 = G_{\mu\nu}d\lambda^\mu d\lambda^\nu .
\end{equation}
Along a trajectory $\lambda^\mu(t)$, this gives
\begin{equation}
    \left(\frac{ds}{dt}\right)^2
    =
    G_{\mu\nu}\dot{\lambda}^\mu\dot{\lambda}^\nu .
\end{equation}
Using the relation between the counterdiabatic Hamiltonian and the quantum metric \cite{delCampo2012},
\begin{equation}
    \Vert H_\mathrm{CD}\Vert^2
    =
    \hbar^2
    G_{\mu\nu}\dot{\lambda}^\mu\dot{\lambda}^\nu ,
\end{equation}
we obtain
\begin{equation}
    \Vert H_\mathrm{CD}\Vert
    =
    \hbar \frac{ds}{dt}.
\end{equation}

The geometric length of the path is
\begin{equation}
    \ell = \int_0^T dt \,
    \sqrt{
    G_{\mu\nu}\dot{\lambda}^\mu\dot{\lambda}^\nu
    }.
\end{equation}
For fixed endpoints, the minimizing path is a geodesic for which the quantity $G_{\mu\nu}\dot{\lambda}^\mu\dot{\lambda}^\nu$ is constant. In that case,
\begin{equation}
    \frac{ds}{dt}=\frac{\ell_\mathrm{geo}}{T}.
\end{equation}
Therefore,
\begin{equation}
    \Vert H_\mathrm{CD}\Vert
    =
    \hbar \frac{\ell_\mathrm{geo}}{T},
\end{equation}
which is time independent.

Equivalently, in Riemann normal coordinates centered at the initial point, geodesics are represented by straight lines,
\begin{equation}
    \lambda'^\mu(t)
    =
    \lambda'^\mu(0)
    +
    \frac{
    \lambda'^\mu(T)-\lambda'^\mu(0)
    }{T}t .
\end{equation}
At the initial point, in the normal coordinates, $G_{\mu\nu}=\delta_{\mu\nu}$. In that case,
\begin{align}
    \Vert H_\mathrm{CD}\Vert
    &=
    \hbar
    \sqrt{
    G_{\mu\nu}(\lambda')\dot{\lambda}'^\mu\dot{\lambda}'^\nu
    }\\
    &= \hbar\frac{\sqrt{\delta_{\mu\nu}[\lambda'^\mu(T) - \lambda'^\mu(0)][\lambda'^\nu(T) - \lambda'^\nu(0)]} }{T},
\end{align}
which is true for the whole trajectory in the case of a one-dimensional geodesic as studied in this article.
\\
\subsection*{Case of three control directions}\label{app:3HCD}

Suppose that one has a two-level Hamiltonian with controls on all directions
\begin{equation}
    H(t) = x(t)\,\sigma_x +y(t)\,\sigma_y+z(t)\, \sigma_z ,
\end{equation}
which we write as $H(t)=\mathbf{h}\cdot\mathbf{\sigma} $ for $\mathbf{h}=(x(t),y(t),z(t))$. The associated counterdiabatic Hamiltonian is given by \cite{Berry_2009}
\begin{align}
    H_\mathrm{CD}(t) &=\hbar \frac{\mathbf{h}\times \dot{\mathbf{h}}}{2\vert\mathbf{h}\vert^2}\cdot \mathbf{\sigma}\\
    &=\Omega_x(t)\,\sigma_x+\Omega_y(t)\,\sigma_y +\Omega_z(t)\,\sigma_z
\end{align}
For a generic trajectory with $x(t),y(t),z(t)\neq0$, the counterdiabatic Hamiltonian has non-vanishing components along all three Pauli directions. In that case, we can only fix the norm of $H_\mathrm{CD}$ with constant counterdiabatic driving:
\begin{equation}\label{condition}
    \Vert H_\mathrm{CD}\Vert^2 = 2(\Omega_x^2(t)+\Omega_y^2(t)+\Omega_z^2(t)) = \hbar ^2\frac{\ell_{geo}^2}{T^2} .
\end{equation}
Even if the norm is fixed, the direction is not and the elements of $H_\mathrm{CD}$ still depend on time. But if one does not initially have control on $\sigma_y$ (i.e. $y(t)=0$), then only $\Omega_y(t)$ is non-zero and is fixed by Eq.(\ref{condition}). In that case, the counterdiabatic Hamiltonian is time-independent:
\begin{equation}
    H_\mathrm{CD} = \Omega_y\,\sigma_y.
\end{equation}

\end{document}